\documentclass{aa}
\usepackage{graphicx,amssymb}

\def\mes{M{\'e}sz{\'a}ros}

\begin{document}

\title{Polarization evolution of the GRB\,020405 
afterglow\thanks{Based on observations made with ESO telescopes at the
Paranal Observatory under programme Id 69.D-0461.}}
\titlerunning{Polarization of GRB\,020405}

\author{S. Covino\inst{1}, D. Malesani\inst{1}, G. Ghisellini\inst{1}, 
D. Lazzati\inst{2}, S. di Serego Alighieri\inst{3}, M. Stefanon\inst{1},
A. Cimatti\inst{3}, M. Della Valle\inst{3}, F. Fiore\inst{4},
P. Goldoni\inst{5}, N. Kawai\inst{6}, G.L. Israel\inst{4}, E. Le
Floc'h\inst{5}, I.F. Mirabel\inst{5,7}, G. Ricker\inst{8},
P. Saracco\inst{1}, L. Stella\inst{4}, G. Tagliaferri\inst{1},
F.M. Zerbi\inst{1}}

\authorrunning{Covino et al.}

\offprints{S. Covino}
\institute{INAF -- Osservatorio Astronomico di Brera, via E. Bianchi 46, 
23807 Merate (LC), Italy.
\and
Institute of Astronomy, University of Cambridge, Madingley Road, CB3
0HA Cambridge, England.
\and
INAF -- Osservatorio Astrofisico di Arcetri, Largo E. Fermi 5, 50125 Firenze,
Italy.
\and
INAF -- Osservatorio Astronomico di Roma, via Frascati 33, Monteporzio Catone
(Roma), Italy.
\and
Service d'Astrophysique, C.E. Saclay, 91191 Gif--sur--Yvette Cedex,
France.
\and
Instituto de Astronom{\'\i}a y F{\'\i}sica del Espacio/CONICET, C.C. 67,
Suc. 28, Buenos Aires, Argentina.
\and
Department of Physics, Tokyo Institute of Technology, 2-12-1 Ookayama, 
Meguroku, Tokyo 152-8551, Japan.
\and
Center for Space Research, Massachusetts Institute of Technology,
Cambridge, Massachusetts 02139--4307, USA.}

\date{}

\abstract{
Polarization measurements for the optical counterpart to GRB\,020405 are
presented and discussed. Our observations were performed with the VLT--UT3
(Melipal) during the second and third night after the gamma--ray burst
discovery. The polarization degree (and the position angle) appears to be
constant between our two observations at a level around $1.5\div2\,\%$. The
polarization can be intrinsic but it is not possible to unambiguously exclude
that a substantial fraction of it is induced by dust in the host galaxy.
}

\maketitle

\section{Introduction}
\label{sec:int}

Polarimetric observations are a unique tool to single out different physical
processes. In the context of gamma-ray burst (GRB) afterglow emission, some
degree of polarization is expected to emerge in the optical flux as a
signature of synchrotron radiation (\mes\ \& Rees \cite{MR97}). The
observation of power-law decaying lightcurves (e.g. Wijers et
al. \cite{WRM97}) and of power-law spectral energy distribution (e.g. Wijers
\& Galama \cite{WG99}; Panaitescu \& Kumar \cite{PK01}) give also further
support to the external shock synchrotron emission scenario.

The first successful polarization measurement was achieved for the optical
afterglow (OA) of GRB\,990510 (Covino et al. \cite{Co99}; Wijers et
al. \cite{WVG99}). Some months later, Rol et al. (\cite{Rol00}) could perform
three distinct observations for GRB\,990712, showing a possible variation in
the polarization degree, but with constant position angle.  More recently,
GRB\,020813 showed definitely a highly significant variation in the
polarization level, again with constant position angle (Barth et
al. \cite{Ba02}; Covino et al. \cite{CMG02}).  Last, for GRB\,021004,
different measurements were performed (Covino et al. \cite{CGM02a},
\cite{CGM02b}; Rol et al. \cite{Rol02}), but the results are still ambiguous
because of the large Galactic-induced polarization. For all these
observations, the polarization degree was always in the range $(0.8 \div
3)\%$.
 
For three further GRBs, GRB\,990123 (Hjorth et al. \cite{Hj99}), GRB\,011211
(Covino et al. \cite{CLM02}) and GRB\,010222 (Bj\"ornsson et al. \cite{Bj02}),
upper limits are again consistent with a maximum value of $\sim 3\%$ ($95\%$
confidence limit).

As a general rule, some degree of asymmetry in the expanding fireball is
necessary to produce some degree of polarized flux.  Gruzinov \& Waxman
(\cite{GW99}) argued that if the magnetic field is globally random but with a
large number of patches where the magnetic field is instead coherent, a
polarization degree up to $\sim 10\%$ is expected, especially at early times.
Ghisellini \& Lazzati (\cite{GL99}) and, independently, Sari (\cite{S99})
considered a geometrical setup in which a beamed fireball is observed slightly
off-axis.  This break of symmetry again results in a significant polarization.
This model also predicts a testable variation of the polarization degree and
position angle associated with the evolution of the afterglow lightcurve.

GRB\,020405 was localized on 2002 April 5 at 00:41:26 UT by the interplanetary
network (IPN) (Hurley et al. \cite{HCF02}). The burst showed a duration of
$\sim 40$~s and therefore belongs to the class of long duration bursts (Hurley
et al. \cite{Hu92}).  The optical counterpart was identified by Price et
al. (\cite{PSA02a}, \cite{PSA02b}) 17.3 hours after the burst as an $R \sim
18.9$ source located at the coordinates $\alpha_{2000} = 13^{\rm h}58^{\rm
m}03\fs12$, $\delta_{2000} = -31\degr22\arcmin22\farcs2$.

VLT observations allowed to determine the redshift of $z=0.695\pm0.005$
(Masetti et al. \cite{MPP02}) and to discover the bright host galaxy (Masetti
et al. \cite{Ma02b}). A new radio source was found at the above coordinates by
the VLA (Berger et al. \cite{BKF02}), with a flux of $0.49$~mJy at $8.46$~GHz.

In addition to those presented here, polarimetric observations were performed
by Masetti (\cite{Ma02}) with the VLT and by Bersier et al. (\cite{Be02}) with
the Multiple Mirror Telescope, beginning $1.2$ and $1.3$ days after the GRB,
respectively. Even if these two measurements were almost simultaneous, their
results are in remarkable contrast. The first group found a level of
polarization $P = (1.5\pm0.4)\%$ (hereafter 1-$\sigma$ uncertainties are
reported) with position angle $\vartheta = (172 \pm 8)\degr$, similar to
other GRBs, while the second group reported the unprecedented high value $P =
(9.89\pm0.13)\%$ at $\vartheta = (179.9\pm3.8)\degr$.  We note however that
the results of both groups are not yet published in a refereed journal.

\section{Data and analysis}

\begin{figure*}[ht]
\includegraphics[width=\columnwidth,keepaspectratio]{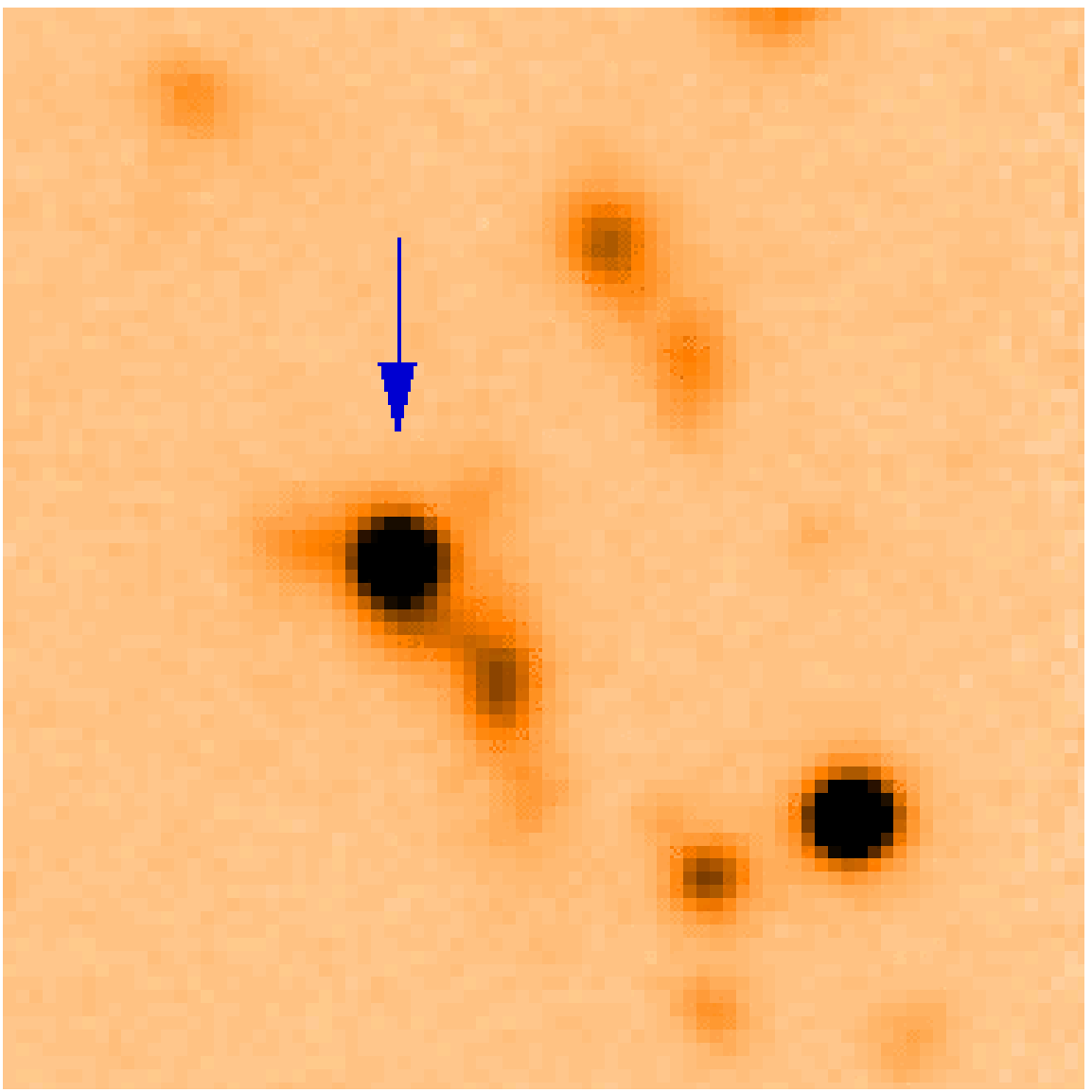}\hfill%
\includegraphics[width=\columnwidth,keepaspectratio]{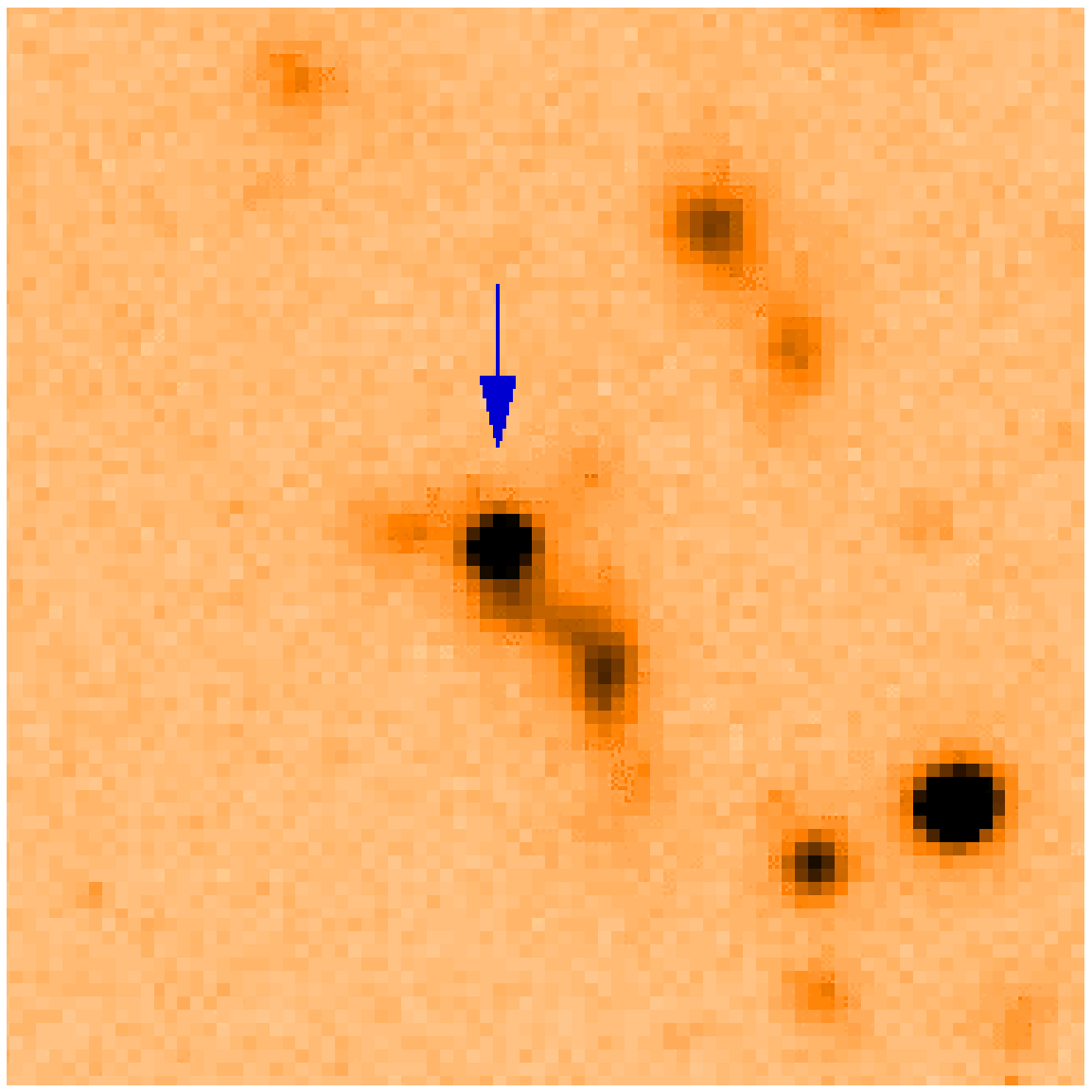}
\caption{
The optical afterglow of GRB\,020405 superimposed to the host galaxy during
run~1 (left) and run~2 (right). Pictures were obtained composing the
polarization images in the Bessel $V$-band filter. Box size is about
$18\arcsec \times 18\arcsec$; North is up and East is left.  The OT is marked
by an arrow.
}
\label{fig:OA}
\end{figure*}

Our observations of GRB\,020405 were obtained with the ESO's VLT--UT3
(Melipal), equipped with the Focal Reducer/low dispersion Spectrometer
(FORS\,1) and a Bessel filter $V$ in the imaging polarimetry mode.  Our first
observation (hereafter run 1) started on April 7, 03:33 UT ($2.1$~days after
the burst) and lasted $\sim 3$~hours.  At the beginning of this observation
the $V$ magnitude was $21.82 \pm 0.02$, with respect to the \mbox{USNO--A2.0}
stars 0525\_16813005 and 0525\_16815468 (Simoncelli et al. \cite{Si02}).  Our
second observation (run~2) was performed during the following night, starting
April 8, 4:01 UT ($3.2$~days after the burst), and lasted $\sim 3.5$~hours.
The $V$ magnitude of the OA was $22.45 \pm 0.05$ again with respect to the two
above reported stars%
\footnote{Note a difference by $\sim 0.4$~mag with respect
to our previous measurement (Covino et al. \cite{CGS02}), due to preliminary
calibration to \mbox{USNO--A2.0} magnitudes.}.
Observations were performed under good seeing conditions ($0.5\arcsec -
0.9\arcsec$) in standard resolution mode with a scale of $0.2\arcsec$/pixel
(Fig.~\ref{fig:OA}).

Standard stars were also observed. One polarized, Hiltner~652, in order to fix
the offset between the polarization and the instrumental angles, and one
non-polarized, \mbox{WD 1615--154}, to estimate the degree of artificial
polarization possibly introduced by the instrument.

The data reduction was carried out with the {\tt Eclipse} package (version
4.2.1, Devillard \cite{D97}). After bias subtraction, non-uniformities were
corrected using flat-fields obtained with the Wollaston prism. The flux of
each point source in the field of view was derived by means of both aperture
and profile fitting photometry by the \mbox{{\tt DAOPHOT II}} package (Stetson
\cite{St87}), as implemented in {\tt ESO}--{\tt MIDAS} (version 01SEP) and the
Graphical Astronomy and Image Analysis ({\tt GAIA}) tools%
\footnote{\texttt{http://star-www.dur.ac.uk/$\sim$pdraper/gaia/gaia.html}\,.}.
For relatively isolated stars the various applied photometric techniques
differ only by a few parts in a thousand. The general procedure followed for
FORS\,1 polarization observation analysis is extensively discussed in Covino et
al. (\cite{Co99}, \cite{CLM02}).

The average polarization of the field stars is low as shown by the
normalized Stokes parameters $Q$ and $U$: $\langle Q \rangle = -0.0021 
\pm 0.0009$ and $\langle U \rangle = 0.0012 \pm 0.0009$, corresponding to 
$P = (0.24 \pm 0.09)$\%.


The degree $P$ and angle $\vartheta$ of polarization are obtained from
the measurements of $Q$ and $U$ for the OA [$P = \sqrt{U^2 + Q^2}$,
$\vartheta = \frac{1}{2}\arctan(U/Q)$] after correcting for the polarization
induced by the instrument or by the local interstellar matter. 
Moreover, for any low
level of polarization ($P/\sigma\le 4$), a correction which takes into
account the bias due to the fact that $P$ is a definite positive
quantity (Wardle \& Kronberg \cite{WK74}) is required. At low
polarization level, the distribution function of $P$ (and of
$\vartheta$, the polarization angle) is no longer normal and that of
$P$ becomes skewed (Clarke et al. \cite{Cl83}; Simmons \& Stewart
\cite{SS85}; Fosbury et al. \cite{Fo93}).

\begin{table}[t]
\caption{Polarization degree $P$ and positional angle $\vartheta$ for the 
optical counterpart to GRB\,020405. Observations were performed with
the VLT--UT3 (Melipal) in the Bessel $V$-band filter.}
\label{tb:pol}
\centering\begin{tabular}{cccccc}
\hline
\bf Run  & \bf UT     &\bf $V$ mag      &\bf $P$ ($\%$)   & \bf $\vartheta$ ($\degr$) \\ \hline
    1    & Apr 7.212  &$21.82 \pm 0.02$ &$1.96 \pm 0.33$  & $154 \pm 5$               \\
    2    & Apr 8.297  &$22.45 \pm 0.05$ &$1.47 \pm 0.43$  & $168 \pm 9$               \\ \hline
\end{tabular}
\end{table}

We then corrected our measurements for this bias (Simmons \& Stewart
\cite{SS85}) and derived the normalized polarization Stokes parameters
for the OA: $Q = 0.0126 \pm 0.0033$ and $U = -0.0150 \pm 0.0033$ for
run~1 and $Q = 0.0137 \pm 0.0043$ and $U = -0.0054 \pm 0.0044$ for
run~2.  From these values of $Q$ and $U$ we have derived the polarization
degree $P$ and positional angle $\vartheta$ for both run~1 and 2, as
reported in Table~\ref{tb:pol}. Monte Carlo simulations confirmed the 
reported values and uncertainties.

\subsection{Host galaxy contamination to photometry}

Fig.~\ref{fig:OA} clearly shows that the OA is superimposed to a
rather bright and extended galaxy (\mbox{$\sim 4\arcsec \times
7\arcsec$} in our VLT images, with some bright knots).  Since the
light of the galaxy is unavoidably mixed with that of the OA, it is
important to estimate the effect of this contamination on the
polarization angle and degree.  If the emission of the galaxy is not
polarized, the net effect is to effectively reduce the degree of
polarization of the OA.  It is easy to show that the
observed polarization degree $P_\mathrm{obs}$ can be corrected to
yield the intrinsic value $P_\mathrm{true}$, if we know the
contributions to the total flux of the galaxy, $F_\mathrm{gal}$, and
of the OA, $F_\mathrm{OA}$:
\begin{equation}\label{eq:Ptrue}
P_\mathrm{true} = \left(1 + \frac{F_\mathrm{gal}}
{F_\mathrm{OA}}\right) P_\mathrm{obs} =
\frac{F_\mathrm{tot}}{F_\mathrm{OA}} P_\mathrm{obs},
\end{equation}
where $F_\mathrm{tot} = F_\mathrm{OA} + F_\mathrm{gal}$.  The
polarization angle is of course not affected, even if the lower value
of $P$ eventually leads to a larger uncertainty.

To estimate the contribution of the galaxy within the point spread function 
(PSF), it is necessary to analyze late-time images, when the flux of the 
afterglow gives only a negligible contribution. For GRB\,020405, only a rough 
$R$ magnitude is reported to date (Bersier et al. \cite{Be02}; see also Price et
al. \cite{Pr02}), suggesting that in the PSF area $V \ge 24$ depending on the color 
of the galaxy (e.g. Fukugita et al. \cite{FSI95}).  

Although an accurate analysis of the late-time image would be required, 
the good seeing conditions in our images make these corrections,
estimated by Eq.~\ref{eq:Ptrue}, essentially negligible.

\section{Discussion}
\label{sec:model}

\begin{figure}
\includegraphics[width=\columnwidth,keepaspectratio]{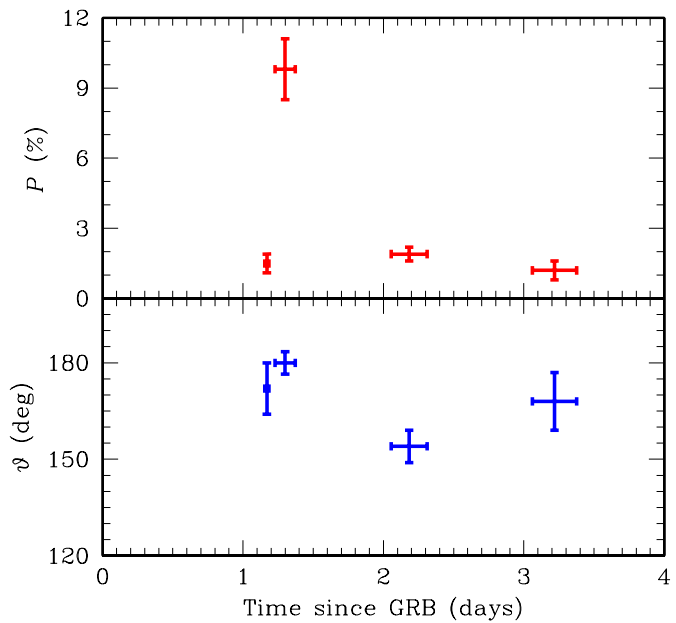}
\caption{Time evolution of the polarization level $P$ and position
angle $\vartheta$. Data for the first night are from Masetti et
al. (\cite{Ma02}) and from Bersier et al. (\cite{Be02}).}
\label{fg:time}
\end{figure}

In Fig.~\ref{fg:time} we show the time evolution of the polarization
level $P$ and angle $\vartheta$ for GRB\,020405, also including the
measurements performed by Bersier et al. (\cite{Be02}) and by Masetti
et al. (\cite{Ma02}). Because of the striking contrast between the
observations of the first night, the question for variability cannot
be firmly settled. However, no significant variation is found by
looking at our data alone (second and third night). Our points are
moreover fully consistent with the one of Masetti et al. (\cite{Ma02}).

A certain amount of (constant) polarization can be introduced by
intervening dust along the line of sight, either in our Galaxy or in
the host. The values reported in Table~\ref{tb:pol} are already
corrected for the (low) Galactic contribution.  If additional dust is
present in the host galaxy, its presence should be revealed through
spectral reddening.
Since the induced polarization should not
be larger than $P_\mathrm{max} = 9\%\,E_{B-V}$ (Serkowski et al. 
\cite{SMF75}), a reddening $E_{B-V} \approx 0.2$ (in the host
frame) would be required to explain our value $P \approx 2\%$. This
transforms into $A_V \sim 0.6 \div 1.1$ depending on the
selective-to-total extinction coefficient $R_V$, that can be higher
than the standard value $\sim 3.1$ in star-forming regions 
(see e.g. Cardelli et al. \cite{CCM89}). 
X--ray data by \textit{Chandra} (Mirabal et al. \cite{MPH02}) indeed
reveal the presence of some material along the line of sight, with
$N_\mathrm{H} = (4.7\pm3.7) \times 10^{21}$~cm$^{-2}$. Assuming a
Galactic dust-to-gas ratio, this corresponds to $A_V = 2.8\pm2.2$
(Predehl \& Schmitt \cite{PS95}). The effect of dust on the polarization degree
can therefore be significant. This shows that the study of
polarization can yield important constraints about the medium
surrounding the GRB progenitor.

In addition to the difficulty of assessing the intrinsic level of
polarization of the OA, interpreting the polarization measurements within
the framework of the proposed models is made difficult by the lack of
a clear break in the power--law decay of the lightcurve.  In fact,
despite some claims of the possible presence of a jet break at early
times ($t_{\rm j} \sim1$~day, Price et al. \cite{Pr02}), the data seem also
compatible with a single power-law up to ten days after the burst
(Masetti et al. \cite{Ma02b}). 

In the framework of the patchy model (Gruzinov \& Waxman \cite{GW99}),
a moderate-high level of polarization is expected. The level of
polarization should monotonically decay as a function of time due to
the increase of the visible surface of the fireball (and therefore to
the increased number of visible patches). The position angle of the 
polarization vector should fluctuate randomly. 
%
Since the polarization predicted in this model is $P \sim
60\%\,/\sqrt{N}$, the inferred number of patches is $N \sim 1000$.

In the case of collimated fireballs, Ghisellini \& Lazzati (\cite {GL99})
and Sari (\cite{S99}) proposed a model in which the polarized fraction
has a more complex behaviour, with two peaks separated by a moment of
null polarization that roughly coincides with the break time of the
total flux lightcurve. Lacking a robust detection of a jet break
and given the limited number of measurements, only a qualitative 
comparison can be performed. Again, the measurement
of Bersier et al. (\cite{Be02}) cannot be reconciled with the model in
any case and, if real, should be ascribed to some still unknown effect
(see Bersier et al. \cite{Be02} for a comprehensive discussion).

In the case of a late time break ($t_{\rm j}>10$~d), our measurements 
can be interpreted to belong to the first peak of the polarization curve 
(see Fig. 4 in Ghisellini \& Lazzati \cite{GL99}), with the moderate decay 
of the polarization being an indication that the break time is approaching. 
If the break were at early times ($t_{\rm j}\le1$~d; see Price et al. \cite{Pr02}),
the absence of a rotation of $90\degr$ of the position angle that 
is predicted between the first and the second peak in the polarization time 
evolution (e.g. Ghisellini \& Lazzati \cite{GL99}, Sari \cite{S99}) would point 
either to a rapidly sideways expanding jet (Sari \cite{S99}) or to a structured 
jet (Rossi et al. \cite{Ro02}, \cite{Rop02}).

\begin{acknowledgements}
We thank the ESO--Paranal staff for the reliable support, and the
referee, Johan Fynbo, for his rapid reply. DM thanks the Italian MIUR
for financial support.
\end{acknowledgements}

\end{document}